\documentclass[mathleft]{an}
\usepackage[latin1]{inputenc}
\usepackage{epsfig}
\usepackage[dvips]{rotating}
\usepackage{graphicx}
\usepackage{times}
\usepackage{makeidx}
\usepackage{fancyheadings}
\usepackage{amsmath}
\usepackage{amssymb}
\usepackage{multirow}

\begin{document}

\DOI{...}

\title{Optimising Optimal Image Subtraction}

\author{Holger Israel\inst{1,2}\fnmsep\thanks{Corresponding author:\newline
  \email{hisrael@astro.physik.uni-goettingen.de}\newline}
\and Frederic V. Hessman\inst{1}
\and Sonja Schuh\inst{1}}

\titlerunning{Optimising Optimal Image Subtraction}
\authorrunning{H. Israel, F.V. Hessman \& S. Schuh}
\institute{Institut f\"ur Astrophysik G\"ottingen, Universit\"at G\"ottingen,
Friedrich-Hund-Platz~1, 37077~G\"ottingen, Germany
\and Argelander-Institut f\"ur Astronomie, Universit\"at Bonn, 
Auf dem H\"ugel~71, 53121~Bonn,
Germany}

\received{...}
\accepted{...}

\keywords{methods: data analysis -- techniques: image processing -- techniques:
photometric}

\abstract{Difference imaging is a technique for obtaining precise relative 
photometry of variable sources in crowded stellar fields and, as such, 
constitutes a crucial part of the data reduction pipeline in surveys for 
microlensing events or transiting extrasolar planets. The \textit{Optimal 
Image Subtraction} (OIS) algorithm of Alard \& Lupton (1998) permits
the accurate differencing of images by determining convolution kernels which, 
when applied to reference images with particularly good seeing and 
signal-to-noise $(S/N)$, provide excellent matches to the point-spread 
functions (PSF) in other images of the time series to be analysed.  
The convolution kernels are built as linear combinations of a set of basis 
functions, conventionally bivariate Gaussians modulated by polynomials. 
The kernel parameters, mainly the widths and maximal degrees of the basis 
function model, must be supplied by the user.
Ideally, the parameters should be matched to the PSF, pixel-sampling, and 
$S/N$ of the data set or individual images to be analysed.
We have studied the dependence of the reduction outcome as a function of the 
kernel parameters using our new implementation of OIS within the IDL-based 
\textsc{Tripp} package.
From the analysis of noise-free PSF simulations of both single objects and 
crowded fields, as well as the test images in the \textit{Isis} OIS 
software package, 
we derive qualitative and quantitative relations between the kernel parameters 
and the success of the subtraction as a function of the PSF widths and sampling
in reference and data images and compare the results to those of other 
implementations found in the literature. On the basis of these simulations, we 
provide recommended parameters for data sets with different $S/N$ and 
sampling.}

\maketitle

\section{Introduction}

Many astrophysical experiments rely on the precise determination of lightcurves
from sources which are either weak, weakly variable, and/or situated in densely
populated backgrounds.
Prominent examples are the detection of extrasolar planets via the transit 
technique, gravitational microlensing, and supernova searches.
As these events are intrinsically rare, dedicated large-scale surveys have only
become feasible due to automated reduction and photometry of large amounts of 
CCD data. An overview of microlensing surveys (\textsc{Ogle}, \textsc{Eros},
\textsc{Macho}, \textsc{Moa}, and \textsc{Planets}) is given in Dominik
\textit{et al.} (2002) and the references cited therein.

Whenever stellar point spread functions (PSFs) overlap, the determination of 
the background correction for aperture photometry is difficult at best and 
heavy blending of stellar profiles renders the assignment of flux to one source
or another ambiguous or even impossible.
Iterative deblending methods of different kinds have been implemented and are 
in wide use:  photometry packages like \textsc{Daophot} (Stetson 1987) or 
\textsc{SExtractor} (Bertin \& Arnouts 1996) succeed in obtaining a high level 
of precision. 
Nevertheless, their accuracy decreases as the degree of blending increases and 
in the densest Galactic fields, it is impossible to deblend without additional
information.

If the source is variable and the background is (roughly) constant, then 
changes in brightness -- if not the absolute brightness -- can be measured if 
one can successfully subtract the non-variable background by taking the 
difference between the images, corrected for any differences in scale and 
seeing; such an analysis is called \textit{Difference Imaging}.
If the images of a time series are compared against a reference image taken 
from the same series, the difference images obtained by adequately subtracting 
the reference image should be empty of any signal except for a few variable 
objects protruding from the background as positive or negative brightness 
variations. 
Their flux relative to the value defined by the reference can then be measured 
by more classical aperture photometry.
For difference imaging to work, the  PSFs in the reference image and in data 
image thus have to be matched exactly.
Gould (1996) and Tomaney \& Crotts (1996) first attempted difference imaging by
adapting the data image to a reference given by the broadest PSF.  
As this deconvolution method deteriorates image quality, it worked only with 
the highest $S/N$ data. 
The nonlinear PSF fitting  introduced by Kochanski, Tyson \& Fischer (1996) was
more robust but numerically time-consuming. 

To date, the most successful difference imaging algorithm is \textsc{Optimal
Image Subtraction} (OIS) first suggested by Alard \& Lupton (1998; 
hereafter AL98) and implemented in their \textsc{Isis} pipeline. 
Starting with Alard (1999), it has successfully been applied in many different 
kinds of photometric surveys:
microlensing campaigns (Alcock \textit{et al.}~1999); surveys for 
variable stars (Olech \textit{et al.} 1999); supernova searches (Mattila \&
Meikle 2001); transit planet searches (Mall\'en-Onalas {et al.} 2003); etc.

Optimal Image Subtraction determines convolution kernels which transform 
reference images 
into the data images via a linear least-squares fit to a pre-defined set of 
basis functions, cleverly avoiding the problems of non-linear fits to the 
parameters of a particular PSF form.
The reference image is either that with the \textit{best} seeing (and $S/N$) or
a coaddition of several good images. 
Because of its linearity, OIS permits the processing of whole images and 
hence uses all available information.

Although the convolution kernels are linear combination of 
pre-defined basis functions, there are still ``external'' parameters which have
to be supplied before the fitting procedure can be started; these parameters 
are explained in detail in Section \ref{sec:ois}.
We investigate the dependence of difference image quality on the values of 
these parameters in Sections \ref{sec:dvec} and \ref{sec:bvec} using different 
sets of simulated data and conclude with an outlook in section 
\ref{sec:outlook}.

\section{Optimal Image Subtraction} \label{sec:ois}

\subsection{The OIS algorithm}

Strictly speaking, the algorithm presented by AL98 does not match PSFs but 
whole images: the convolution of the reference image $R(x,y)$ with a suitable 
kernel $K(u,v)$ results in a model image $J(x,y)$ which represents the best 
approximation in the sense of $\chi^{2}$-fitting to the data image $I(x,y)$. 
The background differences $S(x,y)$ between the images are also fitted 
simultaneously:
\begin{equation} \label{eq:ois}
I(x,y)\,\doteq\,J(x,y) = R(x,y) \otimes K(u,v) + S(x,y) .
\end{equation}
While OIS thus does not require any isolated stellar PSF to be retrievable out 
of the blended profiles in the images, it is still more accurate the more 
distinguishable sources the images contain\footnote{In real life, relatively 
isolated stars remain indispensable for image registration in the existing 
photometry pipelines.}. 

The convolution kernel $K(x,y)$ is a linear combination of basis functions 
$B^{(i)}(u,v)$ with $u$ and $v$ denoting the PSF kernel coordinates:
\begin{equation} \label{eq:kcomp}
K(u,v) = \sum_{i=0}^{N_{\mathrm{psf}}-1}{a^{(i)}B^{(i)}(u,v)}\;.
\end{equation}
The model image is then:
\begin{equation}
J(x,y)\!=\!\!\sum_{i=0}^{N_{\mathrm{psf}}-1}{\!a^{(i)}\!
\left(R(x,y)\cdot B^{(i)}(u,v)\right)}\!+\!S(x,y)\;.
\end{equation}
The convolution being linear, the model images can be expressed by the same fit
parameters $(a^{(i)})$ as a linear combination of $N_{\mathrm{psf}}$ kernel and
$N_{\mathrm{bg}}$ background basis images
\begin{equation} \label{eq:jcomp}
J(x,y) = \sum_{i=0}^{N_{\mathrm{psf}}\!+\!N_{\mathrm{bg}}\!-\!1}
{a^{(i)}C^{(i)}(x,y)}
\end{equation}
with a total number of $N\!=\!N_{\mathrm{psf}}+N_{\mathrm{bg}}$ 
free parameters.

Inserting eq. \ref{eq:jcomp} into the definition of the $\chi^{2}$ estimator 
and approximating the pixel count errors $\sigma(x,y)$ to be normally 
distributed, one obtains a linear system of equations, the normal equations:
\begin{equation} \label{eq:normaleq}
\mathsf{c} = \mathsf{M}\cdot\mathsf{a}
\end{equation}
with $\mathsf{a}$ being the vector of $N$ fit parameters. The elements of the
vector $\mathsf{c}$ and matrix $\mathsf{M}$ are defined as follows:
\begin{align}
\mathsf{M}_{ii'} &= \sum_{x,y}{\left(\sigma(x,y)^{-2}\cdot C^{(i')}(x,y)\cdot 
C^{(i)}(x,y)\right)}\;,\label{eq:mat}\\
\mathsf{c}_{i} &= \sum_{x,y}{\left(\sigma(x,y)^{-2}\cdot I(x,y)\cdot 
C^{(i)}(x,y)\right)}\;.\label{eq:vec}
\end{align}

\subsection{Basis functions according to Alard \& Lupton}

There is no mandatory choice for any particular OIS basis functions.
AL98 define their basis functions $B^{(i)}(x,y)$ as Gaussians of $G$ 
different fixed widths $(b_{g})$ multiplied by polynomials in kernel 
coordinates:
\begin{equation} \label{eq:bf}
B^{(i)}(u,v)=\exp{\left(-\frac{u^{2}+v^{2}}{2b_{g}^{2}}\right)}
\cdot u^{j}v^{k}\;.
\end{equation}
For the exponents $j$ and $k$, the relations $0\leq j,k\leq d_{g}$ and 
$0\leq j+k\leq d_{g}$ hold, where $d_{g}$ is the maximal degree of the 
$g$-th basic Gaussian component. 
The multi-index $i\!=\!(g,j,k)$ comprises the indices running over the basic 
Gaussians and the polynomial exponents. 
The widths $b_{g}$  determine how much a PSF's width increases upon convolution
with the respective kernel and are therefore called \textit{broadening 
parameters}. 
Together with the maximal degrees $d_{g}$, the $b_{g}$ are the \textit{external
parameters} to be supplied beforehand. Further adjustments can be made to the 
number $G$ of Gaussian components and the size $M$ of the array 
representing the convolution kernel for computation. 
The background difference between reference and data image is expanded into a 
polynomial in pixel coordinates with a maximal degree $d_{\mathrm{bg}}$ defined
the same way as the $(d_{g})$. 
Thus, the basis images are the following:
\begin{equation} \label{eq:defc}
C^{(i)}(x,y)\!=\!\begin{cases}\!R(x,y)\!\otimes\!\exp{
\!\left(\!\frac{u^{2}+v^{2}}{-2b_{g}^{2}}\!\right)}u^{j}v^{k} &
i\!<\!N_{\mathrm{psf}} \\ x^{j}y^{k}&i\!\geq\!N_{\mathrm{psf}}\;.
\end{cases} \end{equation}

In the framework of an improved algorithm, allowing for a spatial variation of 
the convolution kernel over the chip to model PSF variations, Alard
(2000) redefined the basis functions by subtracting the $i\!=\!0$ function from
every other basis function of nonzero integral. 
Our difference imaging pipeline within the IDL \textsc{Tripp} package (see 
below) follows the latter definition of the basis functions. 
Further independent implementations of OIS in the literature share the 
definition of basis functions from AL98 or Alard (2000):
e.g. Wo{\'z}niak (2000; hereafter W00); Bond \textit{et al.} (2001); 
G\"ossl \& Riffeser (2002; hereafter GR02).

\subsection{Difference Imaging in the \textsc{Tripp} package}

The Time Resolved Image Photometry Package, \textsc{Tripp},
is an \texttt{IDL} data reduction package for the 
automated processing of large CCD time series. 
Up to now, it has mainly been used for differential photometry with a clearly 
defined object of interest.
Difference image analysis, by focusing on the detection of variable sources in 
crowded (and mostly larger) fields, brings somewhat complementary 
specifications into play.
We have added to the \textsc{Tripp} pipeline, as described in Schuh 
\textit{et al.} (2003), an alternative branch of data flow.
New top-level routines have been added for the interpolation of all images to a
common coordinate grid, the actual PSF matching and image subtraction as well
as image coaddition. 
Shared information, e.g. the external parameters, is provided by an adaptation 
of the log file control from the original pipeline. 

Before resampling an image to the reference grid, saturated pixels and pixels
too close to an image edge to use them for convolution are detected and stored
in a bad pixel mask. Pixels in the new grid to which those bad 
pixels in the original grid contribute are flagged in the final bad pixel mask
and left out in the computation of the $\chi^{2}$ function.

The sums in eqs. (\ref{eq:mat}) and (\ref{eq:vec}) are evaluated on rectangular
subframes of the data images, similar to the methods of AL98 and Alard (2000).
In the normal (fast) mode, one single convolution kernel is obtained from
the combined data in all subframes. In order to account for spatial PSF
variation or regions of particular interest, the user can choose to determine
``local'' kernels for some or all of the subframes which then rely only on
local PSF information.

Variable sources are identified by running an adaptation of the 
\textsc{Daophot} \texttt{find} function (Stetson 1987) on a weighted sum of
difference images. 
For these significantly variable sources, 
flux differences to the reference frame are measured by \textsc{Tripp} 
aperture photometry of the difference images. The flux scale in the difference
images is that of the reference frame; thus, difference lightcurves may
be calibrated by aperture photometry of the reference.

The following investigations on the outcome of difference imaging runs using 
different sets of external parameters have been carried out using the
\textsc{Tripp} difference imaging pipeline. In some analyses, detailed below,
simulated images from the  Alard \& Lupton \textsc{Isis} package were 
processed. The consistency between results from \textsc{Tripp} and 
the \textsc{Isis} pipeline (written in \textsc{C}) was assessed by running both
codes with the same parameters over those images. Except for the more elaborate
handling of edges in \textsc{Tripp}, the 
resulting difference images are similar to the level of showing the same noise
pattern from residuals of source subtraction.

\section{The vectors of maximal degrees} \label{sec:dvec}

\subsection{The maximal degrees and number of parameters}

The number of free parameters is determined by the maximal degrees of the 
modifying polynomials for the $G$ Gaussian components of the convolution 
kernel.
With $\Delta(n)=\sum_{k=1}^{n}{k}$ being the $n$-th triangular number, the 
number of parameters associated with a vector $(d_{1},\ldots,d_{G})$ of maximal
degrees is $\sum_{g=1}^{G}{\Delta(d_{g}+1)}$. (This can be verified by 
counting the possible combinations for $j$ and $k$ in eq. (\ref{eq:bf}).) 
The  simultaneous fit of the differential background adds further parameters.
The benefit in $\chi^{2}_{\mathrm{red}}$ of using higher numbers $N$ of basis
functions must be traded against the runtime of the difference imaging code. 
The $N\!\times\!N$ normal equation matrix and $N$-element vector representing 
the linear problem of $\chi^{2}$-fitting lead to a quadratic increase of 
computations with parameter number. 
Thus, the relation between the $\chi^{2}_{\mathrm{red}}$ and $n$ is of 
practical importance.

\subsection{The simulated point spread functions} \label{sec:psf}

The crucial element of OIS is matching one PSF onto the other by applying an 
appropriate convolution kernel. 
In absence of intrinsic variability, a data image containing an isolated 
stellar profile should ideally be reproduced by the model image resulting from 
PSF matching such that the difference image will show zero flux. 
The fit's $\chi^{2}$ therefore is a direct measure of the residuals from PSF 
matching.

Many of our tests on the OIS algorithm have been carried out using simulated 
test data of single objects.  
Both reference and data images for each set contain a single PSF without noise 
in a $64\times 64$ pixel frame. 
Because our investigations are of the influence of external parameters and not 
the quality of OIS on realistic data \textit{per se}, neither intensity and 
position offsets between images nor noise have been introduced in those data. 

While the reference PSF was chosen to be pure Gaussian by construction, the 
data profile function consists of a weighted sum of a Gaussian and a Lorentzian
function with the same center position: in terms of its radial distance $r$, it
is given by
\begin{equation} \label{eq:lorentz20}
\Phi(r) = \left(1\!-\!\zeta\right)\!\cdot\!\exp\left(-\frac{r^{2}}
{2s_{\mathrm{G}}^{2}}\right)+\frac{\zeta}{1+\left(r/s_{\mathrm{L}}\right)^{2}}
\end{equation} 
with $s_{\mathrm{G}}$ and $s_{\mathrm{L}}$ being the width scales of the
Gaussian and Lorentzian component, respectively.
Throughout this article, the \textit{width} of a profile function will be 
defined as the geometric mean of the standard deviations $s$ of fitting
Gaussian $\exp\left(r^{2}/2s^{2}\right)$ along the Cartesian axes.
Adopting $\zeta\!=\!0.2$ is a reasonable representation of a seeing profile
with a central core and weak extended wings; we will refer to this PSF as 
\textsc{lorentz20}. 
The \textsc{lorentz20} widths were chosen to be 
$s_{\mathrm{G}}\!=\!1.2s_{\mathrm{ref}}$ and  
$s_{\mathrm{L}}\!=\!1.5s_{\mathrm{ref}}$, with $s_{\mathrm{ref}}$ the width of 
the Gaussian reference peak. 
This yields a relation $s_{\mathrm{dat}}\!=\!1.22s_{\mathrm{ref}}$ between the 
final widths in both images.

Due to its Lorentzian shape at large centroid distances~$r$, an isolated 
\textsc{lorentz20} PSF can be measured at radii where it would be completely
dominated by noise under realistic circumstances. 
To localise the PSF, we multiplied the \textsc{lorentz20} PSF with an Gaussian 
envelope function in noise-free simulations. It confines the PSF to the size of
the kernel without changing the relevant central part of the profile.

\subsection{The number of free parameters} \label{sec:parnum}

\begin{figure}
   \includegraphics[width=8cm]{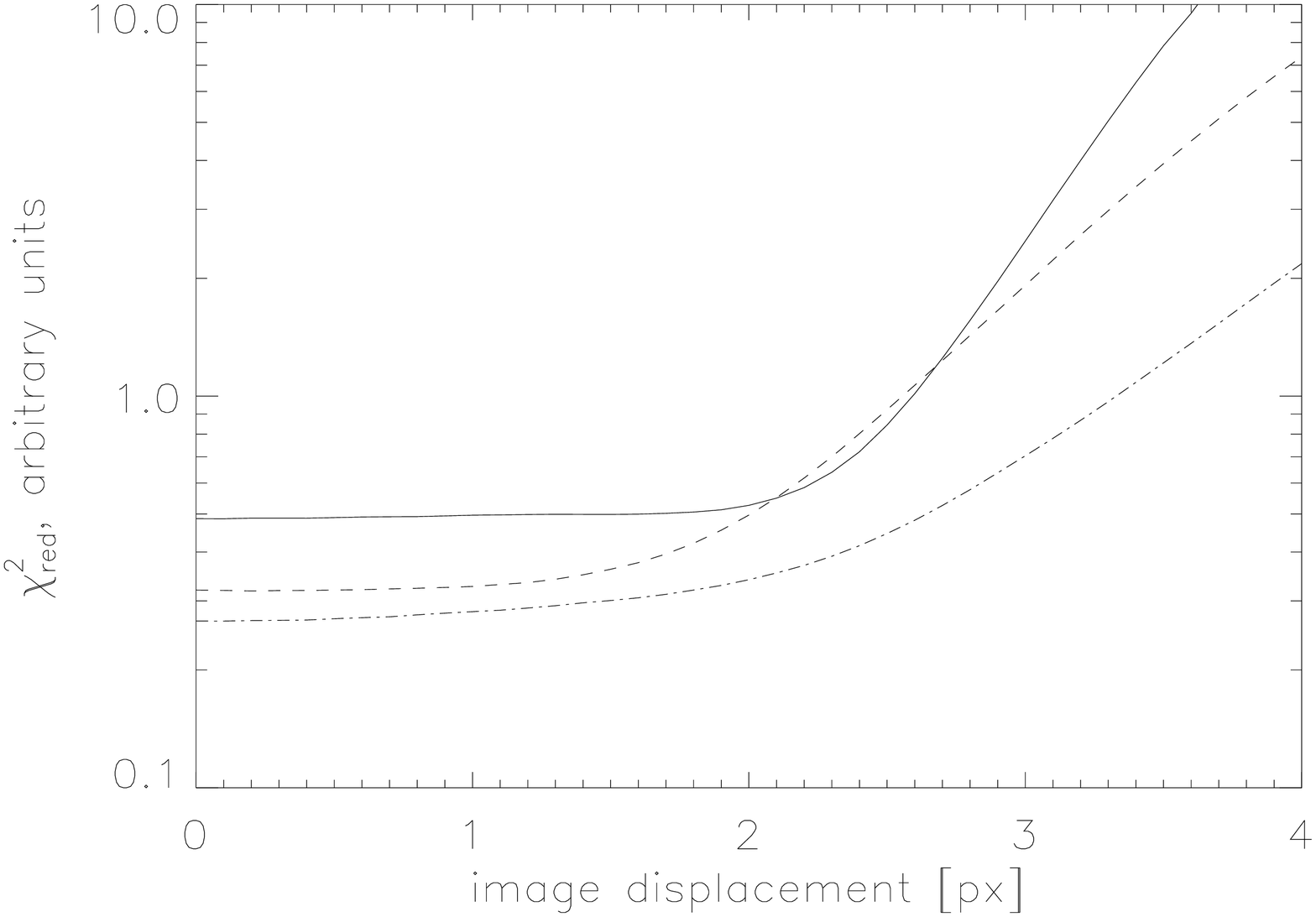} 
   \caption{Dependence of the reduced $\chi^{2}$ on displacements between 
     frames, i.e. residuals of improper image registration. The curves 
     represent the results for different samplings using
     the \textsc{lorentz20} crowded field simulations discussed in 
     sec.~\ref{sec:lor}. The sampling in the data frame was
     \mbox{$s_{\mathrm{dat}}\!=\!1.95$~px} in the continuous curve, 
     \mbox{$s_{\mathrm{dat}}\!=\!2.85$~px} in the dashed curve, and
     \mbox{$s_{\mathrm{dat}}\!=\!3.6$~px} in the dash-dotted one while the 
     sampling is $s_{\mathrm{ref}}\!=\!1.8$~px in the reference frame.}
   \label{fig:odd}
\end{figure}
\begin{figure}
  \includegraphics[width=8cm]{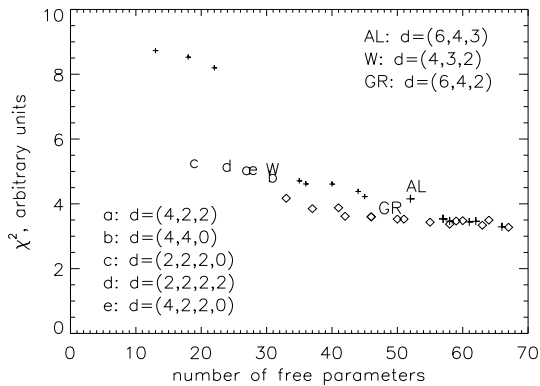}
  \caption[Dependence of the reduced $\chi^{2}$ on the number of free 
parameters.]{The minimal reduced $\chi^{2}$ for the \textsc{lorentz20} data in 
the parameter range $0.5\!\leq\!b_{1}\!\leq\!4.0$~pixels as a function of the 
number $N$ of free parameters given by the maximal degrees assigned to the 
Gaussian kernel components of different width: $G\!=\!3$ (crosses); and 
$G\!=\!4$ (diamonds). Results for the triplets $d\!=\!(4,\,3,\,2)$ (W00), 
$d\!=\!(6,\,4,\,2)$ (GR02) and $d\!=\!(6,\,4,\,3)$ (AL98) are represented by 
the initials of the respective authors.}
  \label{fig:pnum}
\end{figure}

In reductions of perfectly aligned test data, basis function with an odd number
either for $j$ or $k$ contribute nothing to the kernel solution. 
In real data, these odd basis functions permit to process irregular or 
asymmetric PSFs or can shift the PSF centroid and thus 
compensate for any residual misalignment between images.
Test reductions for which an artificial displacement between reference and
data frame had been introduced showed little increase in 
$\chi^{2}_{\mathrm{red}}$ for misalignments $<\!1$~px compared to the case of
perfect image registration (fig.~\ref{fig:odd}).\footnote{
This means that one could -- in principle -- skip the image registration if
the offset between images is sufficiently small and constant.}
Increasing a maximal degree from even to odd is found to decrease 
$\chi^{2}_{\mathrm{red}}$ less than increasing in from odd to even.
For that reason, we will only consider even $d_{g}$.

In the implementations discussed in literature (AL98; W00; GR02), the number 
of Gaussian components is \mbox{$G\!=\!3$}; featuring decreasing maximal 
degrees $d_{g}$ with increasing broadening parameters $b_{g}$.
Kernel components having high polynomial exponents are larger in extent than 
basis functions with lower exponents while the most important part of the
PSF model is the center.
Therefore, in our study, the $d_{g}$ decrease or remain constant with 
increasing $b_{g}$.

We tested the effects of diverse vectors of maximal degrees $(d_{g})$ using 
\textsc{Tripp}. 
Figure \ref{fig:pnum} shows the minima of $\chi^{2}_{\mathrm{red}}$ found in 
the interval 0.5~px~$\leq\!b_{1}\!\leq\!4.0$~px as a function of the number $N$
of free parameters for all vectors of $G\!=\!3$ or \mbox{$G\!=\!4$}.
Results from similar runs using the vectors suggested by AL98, W00, and GR02 
have been added and marked by the authors' initials.\footnote{
Due to the noise-free nature of \textsc{lorentz20} data, the absolute scale of 
$\chi^{2}_{\mathrm{red}}$ is not defined, but can still be used as a relative 
measure of goodness as a function of sample size and parameter number.}

The most noticeable feature in fig. \ref{fig:pnum} are three points at 
significantly higher $\chi^{2}_{\mathrm{red}}$ than most of the others. 
These belong to $d\!=\!(2,\,2,\,0)$ at $N\!=\!13$, $d\!=\!(2,\,2,\,2)$ at 
$N\!=\!18$, and $d\!=\!(4,\,2,\,0)$ at $N\!=\!22$. 
Note that the vector $d\!=\!(2,\,2,\,2,\,0)$ (case \texttt{c} in fig. 
\ref{fig:pnum}) yields a much better $\chi^{2}_{\mathrm{red}}$ at $N\!=\!19$. 
As a rule of thumb, at least 20 parameters are needed for sufficient complexity
of the set of basis function to match point spread functions. 
Beyond this threshold, $\chi^{2}_{\mathrm{red}}$ rather slowly improves with 
increasing $n$.
Between $N\!\approx\!30$ and $N\!\approx\!50$, the use of four widths seems to
yield better results than having $G\!=\!3$, while there is no difference for 
larger $N$.

The efficiency of a certain choice for $(d_{g})$ can be estimated considering
$\chi^{2}_{\mathrm{red}}/N^{2}$. 
The maximal values for this quantity are obtained for those vectors denoted
\texttt{a} to \texttt{e} in fig.~\ref{fig:pnum}. 
This selection also shows that extending  $(d_{g})$ by a maximal degree of zero
(a pure Gaussian basis function) will improve $\chi^{2}_{\mathrm{red}}$, as 
demonstrated by comparing the solutions  $d\!=\!(2,\,2,\,2,\,0)$ vs. 
$d\!=\!(2,\,2,\,0)$ with just one additional parameter. 

The five most efficient vectors have maximal degrees $d_{g}\!\leq\!4$.
This corresponds to $j\!=\!2$, $k\!=\!2$ being the last qualitatively different
distribution of flux enhancements and depressions in the convolution kernel to 
be added with increasing maximal degree. 
{\bf This implies that a further reduction in the PSF matching residuals via a 
higher number of fitted parameters can be most easily obtained by using more 
Gaussian widths and maximal degrees up to four.}
In the rest of this article, we will investigate $d\!=\!(4,\,2,\,2)$  in 
greater detail.

\section{The kernel widths} \label{sec:bvec}

\begin{figure*}
  \includegraphics[width=17cm]{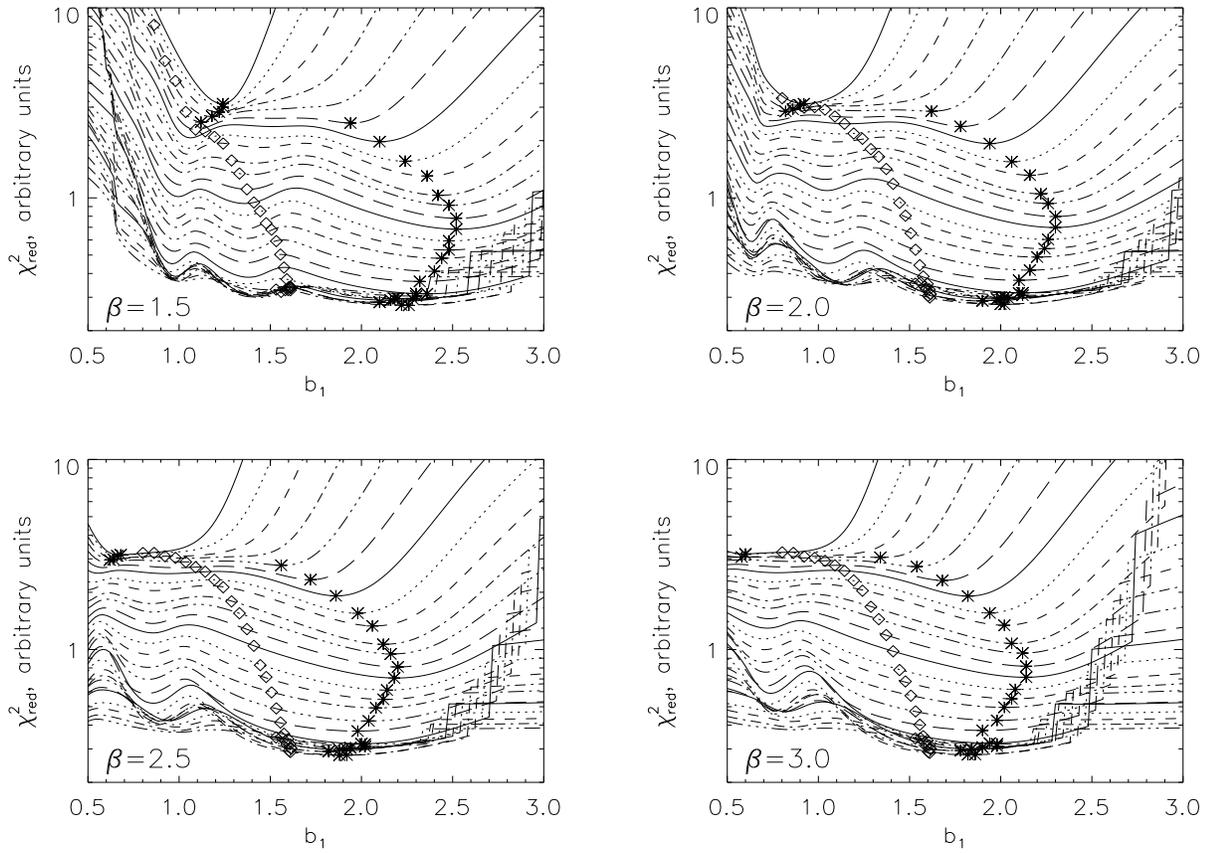}
  \caption{
Reduced $\chi^{2}$ for $d=(4,\,2,\,2)$ as a function of minimum Gaussian kernel
width $b_{1}$ obtained for \textsc{lorentz20} data, kernel spacing parameter
$\beta\!=\!1.5,\,2.0,\,2.5,\,3.0$, and
reference PSF's widths $s_{\mathrm{ref}}$ increasing in steps of 0.1~px from 
$s_{\mathrm{ref}}\!=\!1.2$~px (uppermost line) to $4.0$~px (lowermost at 
$b_{1}\!=\!0.5$~px).
The corresponding data PSF widths runs from $s_{\mathrm{dat}}\!=\!1.44$~px 
(uppermost) to $4.28$~px (lowermost), with step sizes ranging from 0.22 to 
0.08~px. The convolution kernel size was fixed at $m\!=\!19$~px.
Asterisks: positions of the local minima at highest $b_{1}$ for each sampling.
Diamonds: approximate values for $b_{1}$ using $G\!=\!1$, $d\!=\!0$ as in 
eq. (\ref{eq:pyth}).}
  \label{fig:beta}
\end{figure*}

\begin{figure*}
  \includegraphics[width=17cm]{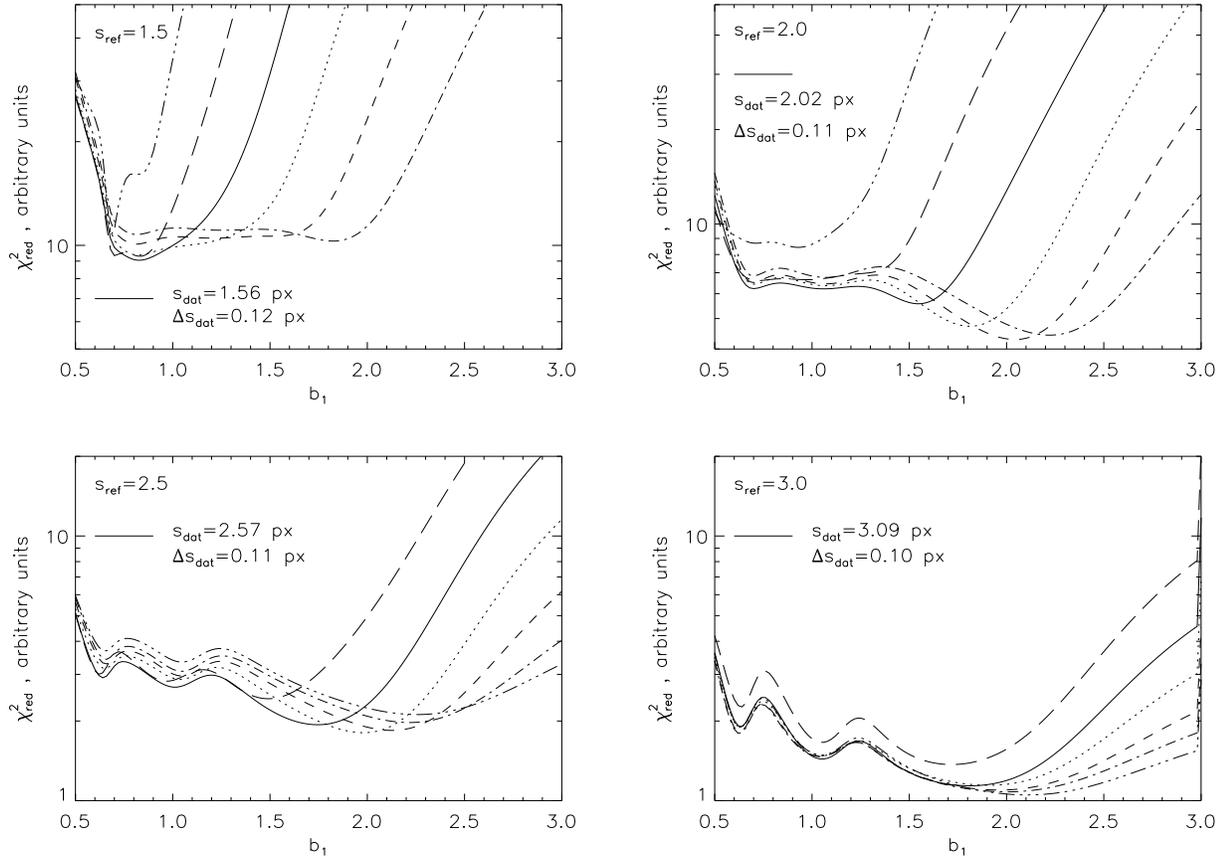}
  \caption{Reduced $\chi^{2}$ as a function of minimum Gaussian kernel width
    $b_{1}$ obtained from reduction
    of \textsc{lorentz20}\,PSFs at four fixed reference image samplings
    $s_{\mathrm{ref}}\!=\!1.5$~px, 2.0~px, 2.5~px, and 3.0~px. The respective
    panels show $\chi^{2}_{\mathrm{red}}(b_{1})$ for a subset of the data
    images from fig.~\ref{fig:beta} with samplings
    $s_{\mathrm{dat}}\!\approx\!s_{\mathrm{ref}}$. A fiducial value of
     $s_{\mathrm{dat}}$ is given in the plots along with the average difference
    between consecutive data image samplings. 
    The kernel spacing parameter is fixed at 
    $\beta\!=\!2.0$ while the maximal degrees and kernel size are set to
    $d=(4,\,2,\,2)$ and $M\!=\!19$~px (as in fig.~\ref{fig:beta}).}
  \label{fig:sref}
\end{figure*}

\subsection{Kernel widths and PSF sampling}

If the number of widths of the kernel components is $G\!=\!3$ or larger, the
parameter space of broadening parameters -- not to mention the whole space for
arbitrary $(d_{g})$ -- becomes complicated.
The different values for the $G$ widths $(b_{g})$ used in the literature have 
been arbitrarily selected by the different authors, based on their own 
experience.

In order to test their selection systematically (and in anticipation of the 
main result of this section result that $\chi^{2}_{\mathrm{red}}$ does not 
depend too sensitively on the $(b_{g})$), we adopted the following system: 
instead of choosing $G$ broadenings independently, we assume they are related 
by a geometrical series such that
\begin{equation} \label{eq:beta}
b_{g} = b_{1}\cdot\beta^{g-1}  ,\quad g=1,\ldots, G\;.
\end{equation}
There remain $b_{1}$, the minimum kernel width, and $\beta$, the kernel spacing
parameter, as independent external parameters. 
The choice of a geometrical series can again be justified by the approximately
Gaussian nature of a PSF with a steep intensity gradient near its centroid and 
wings that fade out super-exponentially into the background.

If the PSF's were perfect Gaussians of widths $s_{\mathrm{ref}}$ in the 
reference image and $s_{\mathrm{dat}}\!>\!s_{\mathrm{ref}}$ in the data peak, 
the convolution kernel mapping the reference onto the data would also be a 
Gaussian of width
\begin{equation} \label{eq:pyth}
b = \sqrt{s_{\mathrm{dat}}^{2}-s_{\mathrm{ref}}^{2}}\;.
\end{equation}
This simple relation can be used as a "first-guess" value of $b_{1}$.
Given $G\!\geq\!1$ Gaussian components and the \textsc{lorentz20} PSF the 
dependence of the optimal value for $b_{1}$ using the geometrical spacing 
(eq.~\ref{eq:beta}) is more complex than the simple relation in 
(eq.~\ref{eq:pyth}). 
For four kernel spacing parameters between $\beta\!=\!1.5$ and 
$\beta\!=\!3.0$, 
values of $\chi^{2}_{\mathrm{red}}$ against $b_{1}$ are presented in 
fig.~\ref{fig:beta}: the graphs show how the locations of local minima in
$\chi^{2}_{\mathrm{red}}$ move with changes in PSF width for
widths of the reference PSFs from 
$s_{\mathrm{ref}}\!=\!1.2$~px to 4.0~px in steps of 0.1~px and associated
data PSFs from $s_{\mathrm{dat}}\!=\!1.44$~px to 4.28~px. 
As a  consequence of limiting the peak's extent by a Gaussian envelope 
function of constant width, there is no fixed relation between 
$s_{\mathrm{ref}}$ and $s_{\mathrm{dat}}$ and the steps in $s_{\mathrm{dat}}$ 
decrease towards wider peaks.

Keeping in mind that PSF shapes look more Gaussian at high $s_{\mathrm{dat}}$ 
in this data set, one can evaluate general properties of the 
$\chi^{2}_{\mathrm{red}}(b_{1})$ curves. 
Smaller PSF widths correspond to seeing conditions more favourable for precise 
photometry. 
On the other hand, on a CCD with its fixed physical pixel size, these peaks are
poorly sampled by their discrete measurements.
Sparsely sampled peaks impose harder constraints on PSF fitting than 
well-sampled ones. 
With increasing PSF width, the convex shape of the $\chi^{2}(b_{1})$ curve has
three minima. 
The global minima in $\chi^{2}_{\mathrm{red}}$, generally located towards the 
highest $b_{1}$ and denoted by asterisks in fig.~\ref{fig:beta},
can lead to a drastic shift in the optimal value for $b_{1}^{\mathrm{opt}}$ 
over a narrow range in  $s_{\mathrm{dat}}$. 
At higher samplings, $b_{1}^{\mathrm{opt}}$, reaches its maximum and 
reapproaches the secondary minimum 
located near $b_{1}\!=\!1.0$~px nearly independent of $s_{\mathrm{dat}}$. 
This effect is not by the changes in data PSF shape and width evoked 
by the envelope function: first guess values for $b_{1}^{\mathrm{opt}}$ derived
from eq. (\ref{eq:pyth}) and marked by diamonds only start to turn at much 
higher $s_{\mathrm{dat}}$ near the bottom of the plots.

Comparing the results for the chosen values of $\beta$, we find the most 
pronounced differences at low samplings. 
For increasing $\beta$, the $\chi^{2}_{\mathrm{red}}$ valley at 
\mbox{$s_{\mathrm{dat}}\!=\!1.44$~px} widens towards lesser values of $b_{1}$
and arrives to be flat for $\beta\!=\!3.0$. 
The observation that larger $\beta$ correspond to lower optimal values for 
$b_{1}$ and tend to result in lower gradients of 
$\chi^{2}_{\mathrm{red}}(b_{1})$  can easily be understood as an effect from 
higher values for $b_{2}$ and $b_{3}$ accompanying a given $b_{1}$. 
This can also be seen from sudden increases of $\chi^{2}_{\mathrm{red}}$ with 
$b_{1}$ which occur whenever a maximum of a basis function contributing 
relevantly to the kernel solution starts to fall onto an edge of the kernel 
array upon incrementing $b_{1}$. 
Such function will then start to act as an additional fitting function to the 
differential background but no longer to the actual PSF matching. 
Figure \ref{fig:beta} shows that edges in $\chi^{2}_{\mathrm{red}}(b_{1})$
appear at smaller $b_{1}$ the higher the value of $\beta$, especially for 
intermediate PSF widths. 
We find the choice of $\beta$ to be uncritical with $\beta\!>\!2.0$, with 
higher $\beta$ allowing a more convenient reduction of narrow PSFs.

Figure~\ref{fig:sref} shows the $\chi^{2}_{\mathrm{red}}$ from similar
reductions at fixed samplings in the reference frame. 
The overall $\chi^{2}_{\mathrm{red}}$ optimum is found at a data 
sampling $s_{\mathrm{dat}}$ slightly larger than $s_{\mathrm{ref}}$. 
Note the ability of the basis system to build image sharpening kernels yielding
a more peaked PSF. 
Our results indicate that a 20\% reduction of PSF width is possible at a 
reasonable $\chi^{2}_{\mathrm{red}}$; a feature 
frequently used for PSF matching of good seeing frames which will be stacked 
to produce a less noisy ''super-reference'' for subsequent difference imaging.

As expected, the location of the minimum moves towards higher $b_{1}$ with
increasing $s_{\mathrm{dat}}$ at a given $s_{\mathrm{ref}}$. 
In addition, we find optimal minimum widths $b_{1}^{\mathrm{opt}}\!=\!0.8$~px 
at $s_{\mathrm{ref}}\!=\!1.5$~px increasing to 
$b_{1}^{\mathrm{opt}}\!=\!2.2$~px at $s_{\mathrm{ref}}\!=\!3.0$~px. 
Shape and width of the $\chi^{2}_{\mathrm{red}}$ valley both depend on sampling
of the reference frame and the sampling difference between reference and data 
image.

The detailed structure of the course of $\chi^{2}(b_{1})$, especially 
$b_{1}^{\mathrm{opt}}$, depends on the individual images to be matched. 
Therefore, after a comparison of several setups, we only discuss the more
robust qualitative features here. 
It should be mentioned, that changing the kernel size $M$ not only reduces 
$\chi^{2}$ but alters the curves for a  given $s_{\mathrm{dat}}$ to a flatter 
shape with respect to $b_{1}$ which resembles that found for broader 
peaks\footnote{
To properly sample the kernel's center element an odd number should be 
chosen for $M$.}. 
We adopted $M\!=\!19$ from AL98.

In further tests including image registration by interpolation, the
minimum value of $\chi^{2}_{\mathrm{red}}$ drastically increases for data
samplings $s_{\mathrm{dat}}\!<1.2$~px, relatively independent of 
$s_{\mathrm{ref}}$.
Interpolation errors dominate the reduction outcome, thus limiting the utility
of OIS for these undersampled data.

\subsection{Crowded field tests using \textsc{isis2} data} \label{sec:isissec}

\begin{figure}
  \includegraphics[width=8cm]{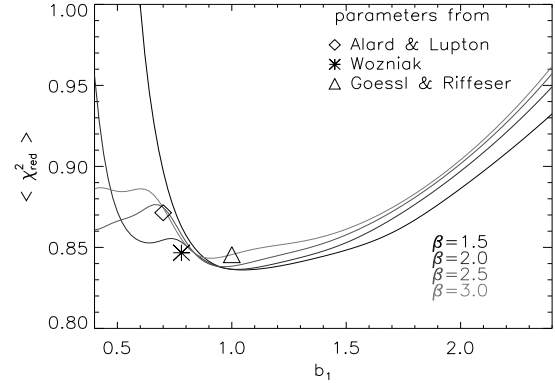}
  \caption[Broadening dependence for \textsc{isis2} reductions. ]{Reduced 
$\chi^{2}$ as a function of $b_{1}$ for a simulated time series consisting of 
every fourth image in the \textsc{isis2} data set and maximal degrees 
$d=(4,\,2,\,2)$. The average over the twelve non-reference images in the series
is denoted by  $\langle\chi^{2}_{\mathrm{red}}\rangle$. 
Curves resulting from \textsc{Tripp}\,difference imaging runs with four 
different $\beta$ factors as well and broadening parameters 
\mbox{$b=(0.7,\,2.0,\,4.0)$} (AL98), $b=(0.78,\,1.35,\,2.34)$ (W00), and 
$b=(1.0,\,3.0,\,9.0)$ (GR02) are plotted.}
  \label{fig:isis}
\end{figure}

In order to assess the parameter dependence of difference imaging quality in a 
more realistic (and easily reproducible) form, we utilised a series of 50 
simulated crowded field observations contained within the \textsc{Isis} package
as test images. 
These data, hereafter referred to as the \textsc{isis2} set, show a nearly 
linear decrease in seeing with index spanning a range from $s\!=\!1.8$\,px to 
$s\!=\!3.2$\,px.
We took the first image as the reference and every fourth of the further ones 
as data images. 
Averaging $\chi^{2}_{\mathrm{red}}$ over the subseries of the latter twelve 
images, the difference imaging quality for different choices of $b_{1}$ and 
$\beta$ are presented in fig. \ref{fig:isis}.

The $\chi^{2}_{\mathrm{red}}(b_{1})$ curves show little variation with the 
tested $\beta$ except for $b_{1}\!<\!0.8$\,px, where 
there is a steep growth in $\chi^{2}_{\mathrm{red}}$ for $\beta\!=\!1.5$ and, 
to a lesser extent, for $\beta\!=\!2.0$ as all three broadenings get small
relative to the PSF difference.
With an optimal value around $b_{1}\!\approx\!1.0$\,px, parameters in the 
interval $0.7~\text{px}\leq b_{1}\leq 1.6~\text{px}$ yield similarly good 
results.
For the higher three values of $\beta$ a secondary minimum at a smaller 
$b_{1}$ exists.

Overplotted in fig. \ref{fig:isis} are results from \textsc{Tripp} 
using $d\!=\!(4,\,2,\,2)$ with broadenings taken from the literature: 
$b\!=\!(0.7,\,2.0,\,4.0)$ (AL98); $b\!=\!(0.78,\,1.35,\,2.34)$ (W00); 
and $b\!=\!(1.0,\,3.0,\,9.0)$ (GR02).
By construction, the latter falls on the $\beta\!=\!3.0$ curve while
W00 employs a geometric spacing of $\beta\!\approx\!\sqrt{3}$. 
The AL98 parameters do not follow a multiplicative relation, but the 
corresponding $\chi^{2}_{\mathrm{red}}$ value is very close to the ones for 
those $\beta$ yielding a comparable vector. 
All of these models come close to the minimum found in our simulations with 
AL98's $b_{1}\!=\!0.7$~px marking the lower limit for advisable $b_{1}$.

With a wider $\chi^{2}_{\mathrm{red}}$ valley and its minimum at 
$b_{1}^{\mathrm{opt}}\!\approx\!2.5$, the \textsc{isis2} reductions differ 
significantly from the \textsc{lorentz} results for a width of the data PSF of 
$s_{\mathrm{dat}}\!\approx\!3.0$~px which is found in the poorer sampled 
\textsc{isis2} images. 
On the other hand, the agreement to the \textsc{lorentz20} model at 
$s_{\mathrm{ref}}\!=\!2.0$~px and $\beta\!=\!2.0$ is fairly good (see 
fig.~\ref{fig:sref}) . 
This can be explained by the fact that, in current implementations of 
difference imaging, \textit{one} set of parameters is chosen for reducing the 
whole time series comprising images obtained under varying observational
circumstances or even collected from several telescopes (as in collaborative 
microlensing surveys). 
Given that situation, the averaging selects for the $(b_{g})$ equalising the
subtraction residuals over the full observed PSF range. 

A closer inspection of the data shows that the overall course of the averaged 
$\chi^{2}_{\mathrm{red}}$ is determined by the images with best seeing next to 
the reference image. 
Their reduction requires slim convolution kernels and thus small $b_{1}$.
In order to obtain the best possible lightcurve, a low 
$\chi^{2}_{\mathrm{red}}$ is especially desirable for images of high quality.
This means applying a small $b_{1}$ to the whole data set.
Small values of $b_{1}\!\approx\!1.0$~px may not be optimal for wide data PSF 
relative to the reference, but the loss in terms of  $\chi^{2}_{\mathrm{red}}$
compared to the minimum is not large. Therefore, the $(b_{g})$ can be
determined by optimising them for the most sensitive data image of best seeing.
As a positive side effect, the success of a certain set of $(b_{g})$ can be 
deduced from a test run on only two images: the reference image and a further 
good seeing exposure.

\subsection{Crowded field tests using \textsc{lorentz20} data} \label{sec:lor}

\begin{figure*}
  \includegraphics[width=17cm]{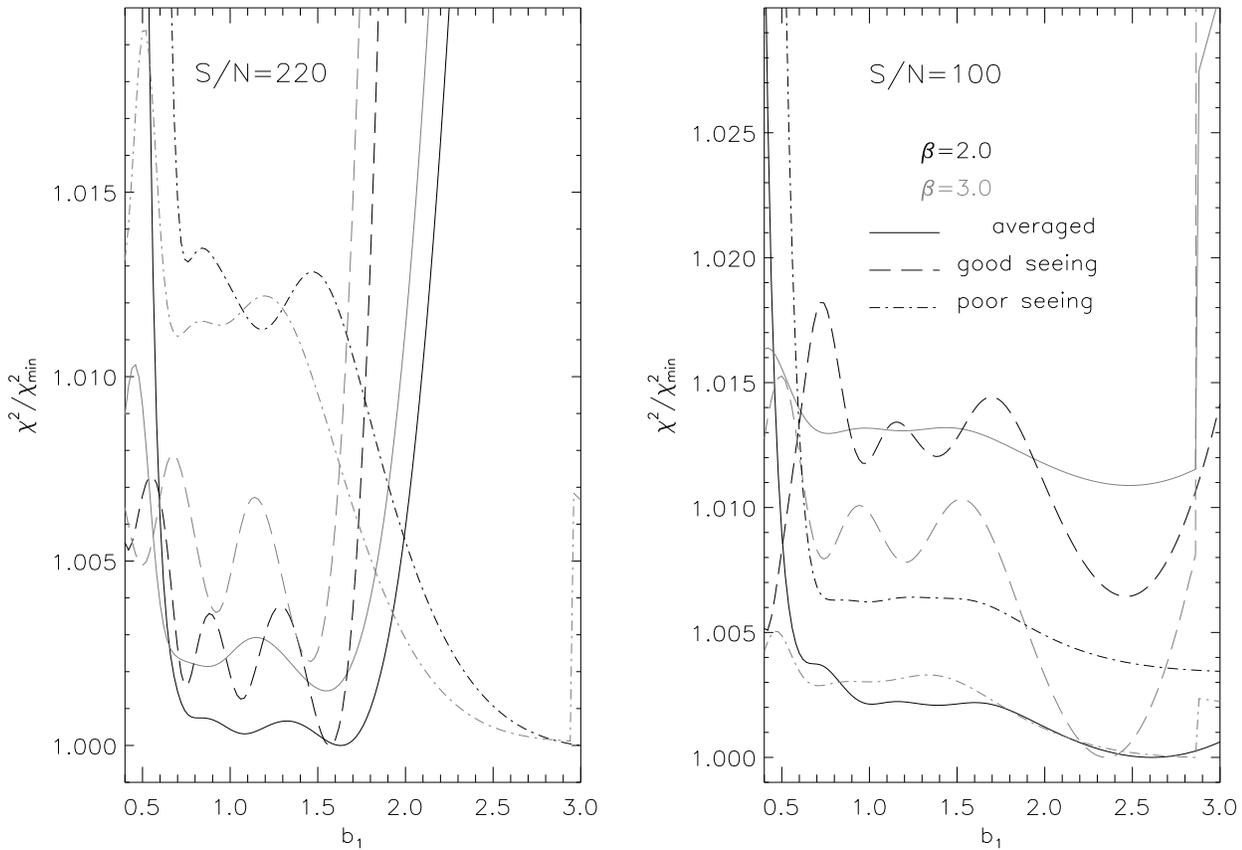}
  \caption{Reduced $\chi^{2}$ as a function of $S/N$ and minimum kernel width
$b_{1}$ for a simulated time series of highly blended \textsc{lorentz20} PSF's.
Shown are $\chi^{2}_{\mathrm{red}}$ averaged over twelve images with linearly 
increasing PSF widths (solid curves) as well as for the best (long dashes) and 
worst (dash-dot) seeing data images.
Note that, with decreasing $S/N$, the differences between reductions of good 
and poor sampling images are also reduced. 
Adding further noise increasingly flattens out the 
$\chi^{2}_{\mathrm{red}}(b_{1})$ curves.}
  \label{fig:snrhm1}
\end{figure*}

Divergent dependences on $b_{1}$ for \textsc{isis2} and 
\textsc{lor\-entz20} data may partly be caused by the obvious 
differences between \textsc{lorentz20} and \textsc{isis2} profiles: 
\textsc{isis2} PSFs are of purely Gaussian shape and are highly blended, 
while the unblended PSF of our test images have extended Lorentzian wings, 
cut off by an Gaussian envelope.

We investigated the relative importance of these effects in reductions of a 
simulated time series of images featuring \textsc{lorentz20} PSFs at 
about the same level of crowding as in \textsc{isis2}. 
Three series of $1\!+\!12$ images at different $S/N$ ratios were created using 
$s_{\mathrm{dat}}\!=\!1.8$~px in the reference images and an increment of 
$+0.15$~px with each higher index in a series. 
Thus, roughly the same seeing range as in \textsc{isis2} is covered. 
The average $S/N$ integrated over an aperture centered on a relatively isolated
star was set at 220 in the highest quality case and reduced to 100 and 30 for 
the other ones. 

Figure \ref{fig:snrhm1} shows how $\chi^{2}_{\mathrm{red}}$ depends on $b_{1}$ 
for the highest and lowest seeing images out of the twelve data images and the 
averages over the whole series for the \mbox{$S/N\!=\!220$} and $S/N\!=\!100$ 
cases.
Values of $\chi^{2}_{\mathrm{red}}$ were normalised to the minimum for the 
specific $S/N$ and sample subset to be able to present the results for 
different grades of seeing in the same plot. 
The $S/N\!=\!30$ curves resemble their counterparts of intermediate quality 
with features smoothed out by the larger random component.   

In the high $S/N$ data set, the $\chi^{2}_{\mathrm{red}}$ averaged over the 
complete sample follows the best seeing data image curve and with 
$0.7~\text{px}\!\leq\!b_{1}\leq\!1.8~\text{px}$ displays a similar interval of 
nearly optimal minimum kernel widths as their $\beta\!=\!2.0$ and 
$\beta\!=\!3.0$ counterparts from the \textsc{isis2} series 
(fig. \ref{fig:isis}).
The shape of the curves $\chi^{2}_{\mathrm{red}}(b_{1})$ for the best and worst
seeing images reproduces quite nicely the results found at both $\beta$ for 
isolated \textsc{lorentz20} peaks of similar reference and data sampling. 
In particular, for the best seeing results at $\beta\!=\!2.0$, the positions of
the minima agree quite well with the ones for the isolated 
$s_{\mathrm{dat}}\!=\!2.0$~px and $s_{\mathrm{ref}}\!=\!2.02$~px (upper right
panel in fig.~\ref{fig:sref}).

Comparing the heavily blended \textsc{isis2} and \textsc{lorentz20} data allows
us to estimate the influence of PSF shape on the difference imaging parameters.
While pure Gaussian PSFs yield a relatively simple dependence 
$\chi^{2}_{\mathrm{red}}(b_{1})$, the multiple minima found with 
\textsc{lorentz20} data seem related to their more complex PSF structure.
Nevertheless, both data sets have in common the range of usable $b_{1}$.  
Blending itself appears to have only a minor influence on the choice of 
$b_{1}$.

With increasing level of noise, the optimum in $\chi^{2}_{\mathrm{red}}$ for 
the  best seeing data image is found at larger $b_{1}$, while for the poorest 
seeing image it becomes less pronounced. 
These tendencies are probably caused by the less clearly defined PSF in noisy 
images. 
It is unclear whether differences between the $\chi^{2}$ levels for the 
same image at different $\beta$ in the right panel of fig. \ref{fig:snrhm1} 
represent an artefact of the simulation.

\section{Conclusion and outlook} \label{sec:outlook}

\begin{table}
  \caption{Summary of the recommended settings for OIS external parameters.}
  \label{tab:res}
  \begin{center} \begin{tabular}{cccc}\hline
      & Maximal & Minimum & Kernel spacing\\
     Situation & degrees & kernel width & parameter\\
      & $(d_{g})$ & $b_{1}$ & $\beta$\\
      & & (px) & \\ \hline
     General & e.g. & & \\
     Recommen- & $(4,\,2,\,2)$, & $0.7\ldots1.2$ & $1.5\ldots3.0$\\
     dation & $(2,\,2,\,2,\,2)$ & & \\ \hline
     Small seeing & e.g. & & \\
     differences & $(4,\,2,\,2)$, & $\approx0.7$ & $1.5\ldots3.0$\\
     to reference & $(2,\,2,\,2,\,2)$ & & \\ \hline
     Good & e.g. & & \\
     overall & $(4,\,2,\,2)$, & $0.7\ldots1.2$ & $\approx3.0$\\
     seeing & $(2,\,2,\,2,\,2)$ & & \\ \hline
     Improved &  e.g. & & \\
     precision & $(6,\,4,\,2)$, & $0.7\ldots1.2$ & $1.5\ldots3.0$\\
     required & $(4,\,4,\,2,\,2)$ & & \\ \hline
  \end{tabular} \end{center}
\end{table}

Optimal image subtraction following the AL98 setup requires about 
$n\!\gtrsim\!20$ parameters for successful subtraction of constant sources
(fig. \ref{fig:pnum}).
The improvement with increasing number of parameters is marginal; the most 
efficient choices for $(d_{g})$ are given in sec. \ref{sec:parnum}.
A larger number $G$ of principal Gaussian should be favoured over maximal 
degrees $d_{g}\!>\!4$.

The choice of kernel widths mainly depends on the differences in PSF widths 
between reference and data images, with eq. (\ref{eq:pyth}) giving a crude 
estimate. 
Given a sampling of the data image $s_{\mathrm{dat}}\!\gtrsim\!1.5$~px, the 
$\chi^{2}_{\mathrm{red}}$ of the reduction does not depend critically on the 
$(b_{g})$. 
Multiplicative spacing of kernel widths proves to be useful, with 
$1.5\!\leq\beta\leq\!3.0$ nearly equally recommendable. 
Due to a small range in widths, values of $\beta\!<\!1.5$ should not be used.
For small differences between reference and data samples, high values of
$\beta$ yield a wider and thus more comfortable interval of low
$\chi^{2}_{\mathrm{red}}$. 
 
Difference imaging has, up to now, applied a single set of kernel widths
to a whole time series, comprising exposures of different seeing.
Parameters should then be chosen to achieve preferentially good reductions for 
high quality images showing a small sampling difference to the reference. 
As outlined in sec. \ref{sec:isissec}, this demands for minimum kernel widths 
in the interval $0.7~\text{px} <\!b_{1}\!< 1.2$~px able to model minute PSF 
differences. 
At even smaller $b_{1}$, basis function themselves become undersampled leading
to a higher $\chi^{2}_{\mathrm{red}}$. 
As a guideline for the user of OIS, these results are summarised in table 
\ref{tab:res}. 
One can, in principle, go on to define customised $(b_{g})$ for individual 
images or groups of images to further optimise $\chi^{2}_{\mathrm{red}}$.

Efforts towards this direction are likely to be impeded by the difficulty of 
deriving quantitative results on the effects of the parameters. 
The deeper reason is that the AL98 basis functions produce elements in the 
matrix $\mathsf{M}$ and the vector $\mathsf{c}$ from eqs. (\ref{eq:mat}) and 
(\ref{eq:vec}) typically spanning $>\!5$ decades, independent of the input 
images.
Although \textsc{Tripp} makes use of the advantages of singular value 
decomposition, small changes in the image data may in some cases lead to
significantly different $\chi^{2}_{\mathrm{red}}$ values. 
This instability coexists with the general features presented above.
Because this problem arises from large numbers introduced by the multiplication
with polynomials of pixel indices in the \textit{ad hoc} definitions in 
(eq.~\ref{eq:defc}), it should be fruitful to look for alternative 
definitions of basis functions. 
Orthogonal functions --  e.g. the decomposition of the PSF into shapelets 
(Refregier 2003) -- might have the double advantages of minimising the number 
of parameters needed and simultaneously allowing for a wider class of PSF 
corrections to be applied by optimal image subtraction.
The matching of seeing conditions with its subtleties continues to be a 
nontrivial problem which solution holds many scientific and numerical insights.

\acknowledgements
H.I. likes to thank Stefan Dreizler for the motivation and support of the
investigations on difference imaging which resulted in this article.

\end{document}